\documentclass[aps,prl,reprint,superscriptaddress]{revtex4-1}
\usepackage[usenames,dvipsnames]{color}
\usepackage{amsmath,array,amssymb}
\usepackage{graphicx}
\usepackage{hyperref}
\usepackage{textcomp}
\usepackage{braket}
\usepackage{multirow}
\usepackage{indentfirst}
\usepackage[FIGTOPCAP,raggedright,nooneline,bf,footnotesize]{subfigure}
\usepackage{paralist}
\usepackage{overpic}
\usepackage{soul} 
\usepackage{xspace} 

\newcommand{\com}[1]{\textcolor{blue}{{\textsc{\textbf{[#1]}}}}}

\newcommand{\cmt}[1]{}

\renewcommand{\eqref}[1]{Eq.~(\ref{#1})}

\newcommand{\tplus}{$^{3+}$\xspace}

\newcommand{\erlinbo}{Er$^{3+}$:LiNbO$_3$\xspace}

\graphicspath{{img/}}

\begin{document}

\title{Effects of mechanical processing and annealing on optical coherence properties of \erlinbo powders}

\author{Thomas Lutz}
\affiliation{
Institute for Quantum Science and Technology, and Department of Physics \& Astronomy, University of Calgary, Calgary Alberta T2N 1N4, Canada
}

\author{Lucile Veissier}
\altaffiliation{Present address: Laboratoire Aim\'e Cotton, CNRS-UPR 3321, Univ. Paris-Sud, B\^at. 505, F-91405 Orsay Cedex, France}
\affiliation{
Institute for Quantum Science and Technology, and Department of Physics \& Astronomy, University of Calgary, Calgary Alberta T2N 1N4, Canada
}

\author{Charles W. Thiel}
\affiliation{
Department of Physics, Montana State University, Bozeman, MT 59717 USA
}
\author{Philip J. T. Woodburn}
\affiliation{
Department of Physics, Montana State University, Bozeman, MT 59717 USA
}
\author{Rufus L. Cone}
\affiliation{
Department of Physics, Montana State University, Bozeman, MT 59717 USA
}
\author{Paul E. Barclay}
\affiliation{
Institute for Quantum Science and Technology, and Department of Physics \& Astronomy, University of Calgary, Calgary Alberta T2N 1N4, Canada
}
\author{Wolfgang Tittel}
\affiliation{
Institute for Quantum Science and Technology, and Department of Physics \& Astronomy, University of Calgary, Calgary Alberta T2N 1N4, Canada
}

\date{\today}

\begin{abstract}
Optical coherence lifetimes and decoherence processes in erbium-doped lithium niobate (\erlinbo) crystalline powders are investigated for materials that underwent different mechanical and thermal treatments. Several complimentary methods are used to assess the coherence lifetimes for these highly scattering media. Direct intensity or heterodyne detection of two-pulse photon echo techniques was employed for samples with longer coherence lifetimes and larger signal strengths, while time-delayed optical free induction decays were found to work well for shorter coherence lifetimes and weaker signal strengths. Spectral hole burning techniques were also used to characterize samples with very rapid dephasing processes. The results on powders are compared to the properties of a bulk crystal, with observed differences explained by the random orientation of the particles in the powders combined with new decoherence mechanisms introduced by the powder fabrication. Modeling of the coherence decay shows that paramagnetic materials such as \erlinbo that have highly anisotropic interactions with an applied magnetic field can still exhibit long coherence lifetimes and relatively simple decay shapes even for a powder of randomly oriented particles. We find that coherence properties degrade rapidly from mechanical treatment when grinding powders from bulk samples, leading to the appearance of amorphous-like behavior and a broadening of up to three orders of magnitude for the homogeneous linewidth even when low-energy grinding methods are employed. Annealing at high temperatures can improve the properties in some samples, with homogeneous linewidths reduced to less than 10 kHz, approaching the bulk crystal linewidth of 3 kHz under the same experimental conditions.
\end{abstract}


\maketitle

\section{Introduction}
Rare-earth-ion (REI) doped solid-state materials have received much attention for their properties such as long optical coherence lifetimes, broad inhomogeneous linewidths and long population lifetimes that are appealing for a multitude of applications \cite{Sun2005,Menager2001,Lauritzen2011,Lauro2009,Afzelius2009,Saglamyurek2014,thiel_rare-earth-doped_2011}. Erbium-doped lithium niobate (\erlinbo) features a wide, inhomogeneously broadened transition within the C-band of the current telecommunication infrastructure \cite{amin_spectroscopic_1996}. This is desirable for high-bandwidth applications such as quantum memories \cite{saglamyurek_broadband_2011} or classical information processing devices, e.g. pulse ordering that would seamlessly integrate with today's technology. Furthermore, lithium niobate is of particular interest due to its strong electro-optic effects \cite{weis_lithium_1985} that enable amplitude and phase modulation of light, as well as its high refractive index that can be readily modified to enable optical waveguide applications \cite{ruter_characterization_2014,armenise_fabrication_1988,sohler_integrated_1999}. The optical coherence properties of the 1.5 micron transition of bulk \erlinbo crystals have also been studied for optical signal processing and quantum information applications \cite{Sun2005,sun_recent_2002,hastings-simon_controlled_2006,thiel_rare-earth-doped_2011,thiel_optical_2010,Thiel2012}, providing a knowledge base for comparing with any changes in the properties of micro- and nano-structured materials.

For many photonic applications, there is a strong trend towards miniaturization of devices with the ultimate goal of on-chip implementation and integration of many optical components. Nanofabrication tools such as chemical etching \cite{benchabane_highly_2009,courjal_acousto-optically_2010} and focused ion beam milling \cite{lacour_nanostructuring_2005} are already available and numerous nano-scale devices have been fabricated specifically from LiNbO$_3$.  Furthermore, employing LiNbO$_3$ crystalline films with thicknesses of only a few microns can provide advantages for the design of next generation high-speed, broadband electro-optic modulators \cite{gheorma_thin_2000,ramadan_electro-optic_2000} as well as providing a platform for integration of heterogeneous photonic structures and components \cite{robinson_silicon_1993}. As fabrication techniques are refined further, additional improvements may be achieved by employing micron-sized ridge waveguide structures, such as for enhanced optical frequency conversion \cite{hu_lithium_2007}. Nanostructured LiNbO$_3$ is of interest for diverse applications such as nanocrystals for in situ biological imaging \cite{knabe_spontaneous_2012} or nano-wires for directional optical second-harmonic generation \cite{grange_lithium_2009} and enhanced pyroelectric sensors \cite{morozovska_pyroelectric_2010}. These materials are also promising for nanoscale ferroelectric capacitors that can enable high-density non-volatile data storage \cite{rudiger_nanosize_2005}. 

While much progress has been made in recent years on nanofabrication of optical materials, little is known about how fabrication processes affect the spectroscopic properties of resonant impurities in the host, such as rare-earth ions. Many properties can be sensitive to the quality of the host crystal, and even modest variations can critically affect applications that rely on those properties. Consequently, it is important to find ways to maintain those properties in nanoscale structures.

In the specific area of REI-doped powders and nanocrystals, there have been very few optical coherence studies. The initial work in this field focused on the specific case of Eu$^{3+}$-doped into sesquioxide host crystals, where fluorescence detection of spectral hole burning revealed rapid decoherence relative to the bulk crystals due to the introduction of glass-like two-level system (TLS) dynamics in the powders \cite{Flinn1994,Meltzer1997,hong_spectral_1998,feofilov_spectroscopy_2002}. Beyond the sesquioxide powders, nano-crystallites in glass ceramics have also suffered from rapid decoherence processes due primarily to interactions with the surrounding amorphous host matrix \cite{macfarlane_spectral_2001,zheng_dynamical_2006,meltzer_photon_2004}. Following the pioneering demonstrations of photon echo measurements on highly scattering powders \cite{beaudoux_emission_2011}, optical coherence properties approaching those of bulk crystals have recently been demonstrated in Eu:Y$_2$O$_3$ nano powders, confirming that there is significant potential to improve these and other nano-materials \cite{perrot_narrow_2013}. In addition, long optical coherence times have been reported for micro-fabricated structures in Nd:Y$_2$SiO$_5$ crystals, again suggesting that at least some systems and fabrication methods can produce properties suitable for practical applications \cite{zhong_nanophotonic_2015}. Nevertheless, beyond these cases, little is known about optical coherence in nano- and micro-structured crystals or powders.

With this motivation, the focus of the work reported here is to employ high-resolution optical characterization methods, namely two-pulse photon echo (2PE), spectral hole burning (SHB), and free induction decay (FID) techniques, to assess optical coherence lifetimes in micro- and nano-sized powders of Er$^{3+}$-doped lithium niobate. The greatest challenges in carrying out these studies involve detecting optically coherent signals in highly scattering media and the complexity in behavior introduced by the random orientation of crystallites in the powder relative to both the external magnetic field and light polarization, which crucially influence the spectroscopic properties.  In addition, we find that the bulk crystal properties degrade significantly with any mechanical processing used to obtain micro/nanocrystals. Therefore, techniques with a very large measurement dynamic range are necessary.

As a primary goal, our study seeks to probe the thresholds where low-energy mechanical processing methods begin to degrade the optical coherence properties and to what extent the effects might be reversed by high-temperature annealing. In addition to the direct relevance to nano- and micro-fabrication, these results can also reveal insights into potential surface damage introduced by cutting, grinding, and polishing as well as what processing steps may be successful in reducing the resulting surface strain and defects. Another important goal is to demonstrate that powders of highly anisotropic materials may be quantitatively analyzed within the framework of decoherence models developed for bulk single crystals. Finally, this work demonstrates how combinations of 2PE, SHB, and FID measurement techniques allow optical coherence properties to be studied and compared across a range of samples with very different behaviors as described in the following.

The following section of this paper describes the fabrication of our powder samples. We then discuss and compare the results of the different experimental methods to probe the optical coherence lifetimes. The properties of the fabricated powders are then compared with the ones of a bulk crystal, including quantitative modeling of the effects due to the random orientation of crystallites in the powder. We identify the factors limiting the optical coherence lifetimes in small crystals and show that high-temperature annealing to relieve strain induced by the fabrication process can improve those properties; however, in the majority of samples, the bulk crystal properties could not be preserved or fully recovered.

\section{Sample preparation}

Mechanical grinding is a very general, simple, rapid, and flexible approach for fabricating nano-powders from bulk crystalline materials.  In particular, high-energy planetary ball milling has been extensively used to produce monodisperse nano-powders with crystallite sizes down to less than 10 nm \cite{koch_synthesis_1993,sepelak_transformations_2012}. However, it is known that high-energy ball milling can cause significant damage to the crystal structure, with the resulting disorder leading to amorphous-like behavior or even transformations to different crystal phases after extended milling \cite{sepelak_transformations_2012,michel_nanocrystalline_1995}. Since very large internal electric fields can be produced through the piezoelectric and pyroelectric effects in LiNbO$_3$ \cite{weis_lithium_1985}, which are also known to contribute to optical damage processes in the bulk crystals \cite{bravina_low-temperature_2004}, we might expect even greater potential for crystal damage from high-energy milling of this material. High-energy ball milling has been used by a number of groups to produce LiNbO$_3$ nanopowders, mostly undoped, and the resulting materials have been extensively characterized \cite{bork_nmr_1998,pooley_synthesis_2003,heitjans_fast_2004,chadwick_lithium_2005,heitjans_nmr_2007,kar_preparation_2013}. A range of techniques including nuclear magnetic resonance spectroscopy, extended X-ray absorption fine structure spectroscopy, impedance spectroscopy, X-ray diffraction, transmission electron microscopy, differential scanning calorimetry, thermogravimetic analysis, infrared absorption spectroscopy, and Raman spectroscopy have all indicated a significant increase in amorphous behavior in the LiNbO$_3$ crystallites when they are reduced to sizes below 100 nm by high-energy ball milling \cite{bork_nmr_1998,pooley_synthesis_2003,heitjans_fast_2004,chadwick_lithium_2005,heitjans_nmr_2007}.

Very little is known regarding how the extent of crystal damage caused by mechanical fabrication or processing relates to the energy, processing time, and specific method used.  While it is clear that the extended high-energy milling required to produce nanocrystalline powders produces significant damage, it is unknown if the damage is intrinsically linked to the size reduction or whether lower energy grinding methods might better preserve the properties of the bulk crystal. For LiNbO$_3$, some of the same measurements that reveal the large degradation in nanocrystals indicate that microcrystals have nearly the same properties as the bulk crystal \cite{heitjans_nmr_2007}. In our previous SHB studies of Tm$^{3+}$-doped yttrium aluminum garnet powders, we found that high-energy planetary ball milling dramatically affected $^{169}$Tm$^{3+}$ nuclear hyperfine state lifetimes; however, we also found that the lifetimes were still significantly reduced in relatively large microcrystals produced by low-energy ball milling \cite{lutz_effects_2016}. Since we expect crystal defects introduced by grinding to affect optical coherence properties more strongly than nuclear spin lifetimes due to the stronger coupling of the electronic states to the crystal lattice, optical coherence measurements should provide a uniquely sensitive method for probing even very low levels of crystal damage. For the same reasons, we might expect mechanical fabrication methods to be particularly detrimental to material performance in applications that rely on long coherence lifetimes and narrow optical homogeneous linewidths.  Consequently, in this work we only used very slow and low-energy grinding methods in an attempt to minimize the accumulated mechanical strain in the crystallites and also probe the threshold for introducing measurable damage to the crystal structure.

\begin{figure}[t]
\centering
\includegraphics[width=.9\columnwidth]{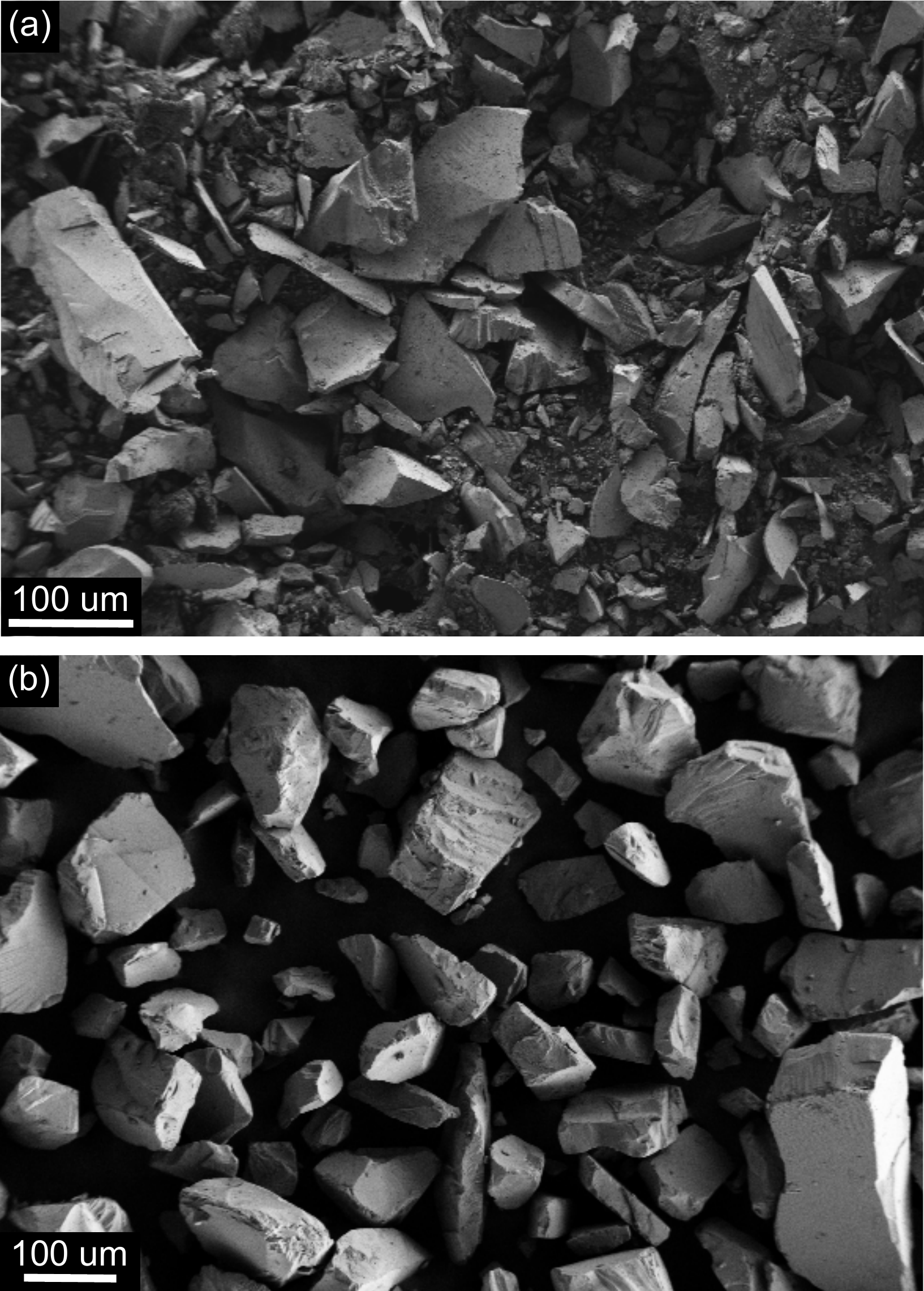}
\caption{Scanning electron microscope images of (a) powder \#1 made from a 0.1\% \erlinbo bulk crystal (SMC) by grinding with a mortar and pestle, (b) powder \#2 ball-milled for only 30 min.}
\label{fig:sempicsunp}
\end{figure}

\begin{figure}[t]
\centering
\includegraphics[width=.9\columnwidth]{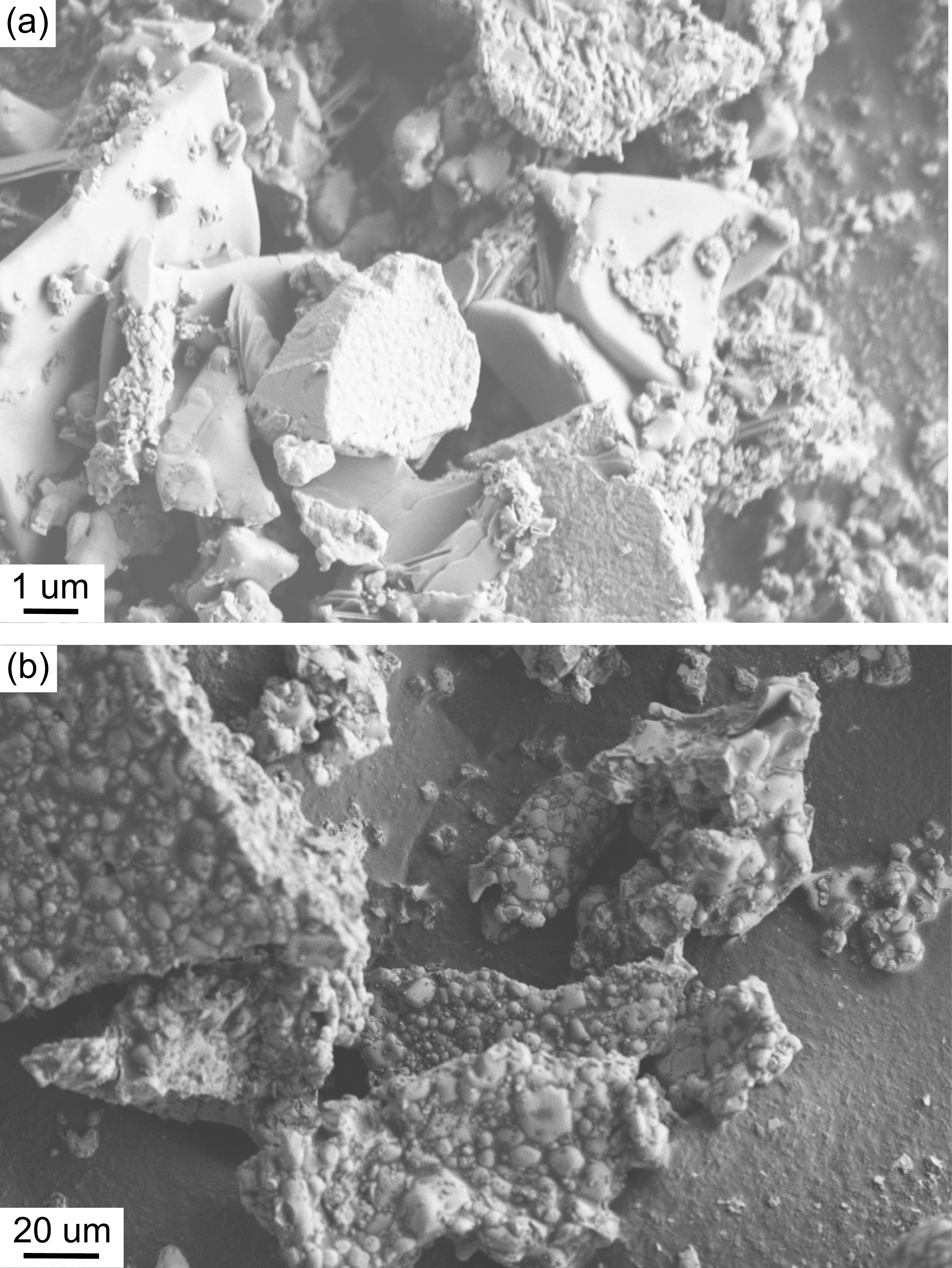}
\caption{Scanning electron microscope images (a) powder \#3 ball-milled and annealed at 900 C, and (b) powder \#6 annealed up to 1100 C. }
\label{fig:sempicsproc}
\end{figure}

As a single crystal reference to compare with powder samples prepared in different ways, we chose a bulk single crystal of 0.1\% \erlinbo with congruent composition, grown by Scientific Materials Corp. (SMC) using the Czochralski method. This crystal has been previously studied by Thiel et al. \cite{thiel_optical_2010,Thiel2012} and showed good coherence properties with homogeneous linewidths as narrow as 3 kHz with applied magnetic fields along the crystal's $c$-axis. The choice of doping concentration is a trade-off between the absorption needed to obtain a reasonable signal-to-noise ratio in powder samples and the additional decoherence created by increased doping due to Er$^{3+}$-Er$^{3+}$ interactions. Starting from a bulk crystal, we fabricated several powder samples, as detailed below. The procedures used to fabricate each of the samples are outlined in Tab.~\ref{tab:fab}. In order to analyze the size and shape distribution of the resulting powders, we performed scanning electron microscope (SEM) imaging, as presented in Fig.~\ref{fig:sempicsunp} and ~\ref{fig:sempicsproc}.

\begin{table}[t]
\centering
\begin{tabular}{|l|l|}
\cline{1-1}
\hline
\multicolumn{1}{|l|}{Sample} &   Method \\
\hline
\hline
bulk   & 0.1 \% \erlinbo (SMC)  \\
\hline
1 \cmt{\com{39}}(ref)  & bulk crushed via mortar and pestle  \\
\hline
2 \cmt{\com{61}}       & \begin{tabular}[c]{@{}l@{}}sample 1 re-crushed with mortar and pestle\\ 
ball milled, 0.5 Hz mill, 30 min
\end{tabular}\\
\hline
3 \cmt{\com{44}}   & \begin{tabular}[c]{@{}l@{}}
sample 1 ball milled, 0.5 Hz mill, 2 hr  \\ 
annealed (1 hr at 900 C and 4 hr at 600 C)
\end{tabular}  \\
\hline
4 \cmt{\com{48}}   & \begin{tabular}[c]{@{}l@{}}
sample 1 ball milled, 0.5 Hz mill, 2 hr \\ 
annealed (4 hr at 1100 C)
\end{tabular}\\
\hline
5 \cmt{\com{49}}   & \begin{tabular}[c]{@{}l@{}}
sample 1 ball milled, 5 Hz mill, 10 hr \\ 
annealed (4 hr at 950 C and then 4 hr at 1100 C)  \\  
\end{tabular}\\
\hline
6 \cmt{\com{50}}    & \begin{tabular}[c]{@{}l@{}}
indiffusion of Er$^{3+}$ into commercial LiNbO$_3$ powder:\\
\ \ \ LiNbO$_3$ +  ErCl$_3$ ball milled, 5 Hz mill, 3 hr \\ 
\ \ \ annealed (4 hr at 1100 C) 
\end{tabular}\\
\hline
\end{tabular}
\caption{Powders characterized in the manuscript and their fabrication methods.}
\label{tab:fab}
\end{table}

An initial reference powder (powder \#1 in Tab.~\ref{tab:fab}) was first fabricated by gentle mechanical crushing and grinding using an alumina ceramic mortar and pestle. This resulted in monocrystalline particles with a wide distribution of sizes ranging from 1 to 100 $\mu$m, as shown in Fig.~\ref{fig:sempicsunp}(a). However, we believe that when we later optically probed the samples the signal mainly originated from the largest particles.

To further reduce the crystallite size of the crushed LiNbO$_3$ powder, we employed two different horizontal tumbling-type low-energy ball mills: a ball mill with a 4-inch diameter HDPE-jar rotating at a relatively slow rate of approximately 5 Hz, and an even slower ball mill rotating at a rate of only approximately 0.5 Hz. Both mills were loaded with a mix of ceramic yttria-stabilized zirconia balls with diameters ranging from 0.3 mm to 1 cm.  The reference powder samples were mixed with ethanol to produce a slurry and then ground in a mill under the conditions listed in Tab.~\ref{tab:fab}. This method was chosen to minimize both the mechanical and thermal energy involved in the grinding. From the powder sample \#1, we produced another sample (powder \#2) by ball-milling for only 30 min at 0.5 Hz. The corresponding SEM image (Fig.~\ref{fig:sempicsunp}(b)), shows very little decrease in particle size demonstrating that our low-energy ball milling method only reduces the particles very slowly by gradual attrition. Nevertheless, even short periods of milling may potentially introduce measurable lattice strain that can be characterized. The next sample fabricated, powder \#3, was produced by ball-milling powder \#1 for 2 hours at 0.5 Hz, resulting in a significant reduction in average particle size. As shown in Fig~\ref{fig:sempicsproc}(a), this sample has particle sizes roughly on the order of  1 $\mu$m after annealing in the furnace.

To investigate the effect of post fabrication thermal treatment, some powders were annealed in alumina crucibles in a tube furnace using a mullite tube providing peak temperatures of up to 1400 C. While thermal annealing can increase defect mobility and relieve crystal strain, LiNbO$_3$ can also be detrimentally affected by annealing at high temperatures due to the potential for out-diffusion of both lithium and oxygen, affecting the optical and electrical properties of the crystal. Past studies have found that thin crystalline films of LiNbO$_3$ are particularly sensitive to high annealing temperatures \cite{wang_influence_2007,budakoti_enhancement_2008}, suggesting that similar effects might be observed for powders. Furthermore, while bulk crystals generally require reducing atmospheres \cite{jhans_optical_1986} or vacuum \cite{dhar_optical_1990} for significant oxygen to diffuse out of the crystal matrix during annealing, we found that in small crystallites the out diffusion occurs even when annealed in an ambient atmosphere, resulting in black powders. To prevent the loss of oxygen from the material, we annealed the powders under a continuous flow of oxygen (ultra-high purity grade), a method that has been shown to inhibit the reduction process for bulk crystals \cite{dhar_optical_1990}, although for thin films there have been some indications that excess oxygen may potentially be absorbed and produce defects \cite{simoes_influence_2004}. To inhibit the out-diffusion of lithium, we employed the well-known technique used for LiNbO$_3$ wafer processing of introducing small amounts of water vapor into the oxygen by bubbling the gas through a column of deionized water \cite{jackel_elimination_1981,forouhar_effects_1984}.

The effects of several different annealing procedures were examined. Powder \#3 was annealed at a low-temperature of around 900 C, which resulted in a growth in particle size of only a factor of  two at most for the micron- and sub-micron-sized particles and only produced weak agglomeration of the powder. The resulting maximum particle size of 2 $\mu$m or less constitutes the finest powder examined in this study as shown in Fig.~\ref{fig:sempicsproc}(a). We also performed higher temperature annealing at 1100 C that tended to produce significant crystallite growth as well as hard agglomeration and fusing of the particles.  Samples \#4, \#5 and \#6  had initial particle sizes of less than 2 $\mu$m after ball-milling but fused into much larger masses around 100 $\mu$m after annealing at 1000 C, as shown in Fig.~\ref{fig:sempicsproc}(b). Powder \#5 also underwent an initial pre-annealing step at 950 C. Better crystal quality, and therefore improved coherence properties should in principle come with this increase in size.

Finally, we fabricated an \erlinbo powder (sample \# 6) via in-diffusion of Er$^{3+}$ ions into a commercial congruent LiNbO$_3$ powder (Sigma Aldrich, 99.9\% purity) with initial crystallite sizes of less than 70 $\mu$m. The LiNbO$_3$ powder was mixed with a 5\% molar concentration of anhydrous ErCl$_3$ powder (ProChem Inc., $>99.9$\% purity) and ethanol to produce a slurry, and the mixture was ball milled for 3 hours. The powder was then allowed to dry in a desiccator, after which it was heated in the tube furnace for 4 hours at 1100 C while under a continuous flow (3 SCFM) of oxygen. The resulting powder was then crushed with a mortar and pestle and washed with deionized water followed by ethanol. Optical spectroscopy of the 1532 nm absorption line verified that significant Er$^{3+}$ had diffused into the LiNbO$_3$ powder. While the amount of Er$^{3+}$ incorporated into the LiNbO$_3$ powder is unknown, the magnitude of the optical absorption (see Tab.~\ref{tab:hb}) and coherence signals was comparable to our other 0.1\% Er powders, suggesting a similar effective doping level for the indiffused powder.

\section{Optical coherence lifetime measurements} 

Key requirements for many applications of REI-doped materials in classical and quantum signal processing are long optical and spin coherence lifetimes. To investigate the effect of powder fabrication on the optical coherence lifetime of the 1.5 $\mu$m telecom-compatible transition of Er$^{3+}$, we used several different measurement techniques.
One issue that must be overcome for powders is a much weaker signal strength. Powders with particle sizes below 10 $\mu$m scatter light very strongly, especially for our samples, since LiNbO$_3$ has a relatively high refractive index of approximately 2.2 at 1.5 $\mu$m wavelengths \cite{schlarb_refractive_1993,zelmon_infrared_1997} and most of our particles have sizes larger than the wavelength of the probing light. The latter causes light to be scattered at every surface of the micro/nano-crystals, randomizing the direction, path length, and polarization of the light, making the transmitted or emitted light difficult to collect. Therefore, the usual characterization methods employed for REI-doped crystals such as 2PE, SHB, or FID are difficult to perform with powders. Previous measurements on powders have mostly employed detection of fluorescence emission to observe SHB \cite{Flinn1994,Macfarlane2004,hong_spectral_1998}, with direct transmission detection of SHB \cite{lutz_effects_2016} and 2PE \cite{beaudoux_emission_2011,perrot_narrow_2013} only recently being successfully applied to powders. In particular, FID measurements on powders have not been previously reported to our knowledge. 

Another consideration is the random orientation of individual crystallites in the powder relative to applied external fields. It is known that Er$^{3+}$ coherence lifetimes strongly depend on the magnetic field orientation due to the anisotropy of the $g$ tensors in both the ground and excited states \cite{bottger_effects_2009}. As a consequence, the coherence lifetimes measured in a randomly orientated powder sample correspond to an average of the bulk lifetime over all possible orientations, complicating the interpretation of the results.

Finally, the degradation of coherence properties in micro/nano-crystals due to new decoherence mechanisms introduced by the powder fabrication process can limit the range of materials that can be studied using a single technique. In some cases, the optical coherence lifetime is drastically reduced, making a direct measurement via 2PEs difficult with standard techniques.

In the work reported here, we employed a combination of three techniques, SHB, 2PE, and FID, to study optical coherence over the wide range behavior encountered in different powder samples. Also, SHB signals were detected in transmission rather than emission to allow information about absorption depths to be determined. For all of these measurements, unless mentioned otherwise, the powder samples were loaded into a 0.4 mm thick space within an unsealed glass cuvette and then cooled to a temperature of 1.6 K in an Oxford Spectromag cryostat using helium exchange gas. A superconducting solenoid provided magnetic fields of up to 6 T for these measurements.  Since most of the incident laser light was diffusely scattered in all directions, it was critical to minimize stray light from reaching the detector and producing a large background signal.  For this purpose, the glass cuvette containing the powder was mounted against a mask with a small hole that only allowed light at the center of the cuvette to be transmitted and with extensive light baffles on all sides to direct any scattered light away from the detector.

\subsection{Two pulse photon echoes}
The most precise method to characterize long coherence lifetimes, as observed in \erlinbo bulk crystals, is the 2PE technique since it is not limited by the linewidth of the excitation laser. For these measurements, a homemade Littman-Metcalf extended-cavity diode laser emitting at 1532 nm was amplified using an erbium-doped fiber amplifier (IPG Photonics Corp EAD-1-C). Two short pulses, separated by a time $t_{12}$, with duration ranging from 30 to 100 ns, optimized for each sample, were generated using an acousto-optic modulator (Crystal Technology Inc. 3165). The light was then focused on the front face of the sample and the transmitted light was detected using an amplified high-speed InGaAs photodetector (New Focus 1811) with no optical gate used between the photodetector and the powder samples. 

In addition to direct detection of the echo intensity, we also performed heterodyne measurements of the echo. For those measurements, after the second excitation pulse, a weak optical local oscillator pulse detuned by 10 MHz was sent through the sample to the detector to ensure it undergoes the same scattering path and to provide better spatial mode matching of the local oscillator and the excitation beam. The envelope of the beat signal between the emitted echo and the local oscillator corresponds to the emitted temporal shape of the photon echo electric field strength. Because of the need to limit the local oscillator pulse intensity to avoid undesired hole burning or heating effects and attenuation due to scattering, the heterodyne signal strength was not significantly larger than the intensity detection.  Nevertheless, the factor of two slower echo decay rate of the electric field observed in the heterodyne measurement, compared to the decay rate of the intensity observed in a regular 2PE, allows shorter coherence lifetimes to be characterized with the heterodyne technique.

\subsubsection{Echo decays in powders of randomly orientated crystallites}

In the simplest 2PE decay measurement, the area of the echo emitted at a time $2 \, t_{12}$ after the first excitation pulse follows an exponential decay determined by the coherence lifetime. In \erlinbo, as for many other Er$^{3+}$ materials, spectral diffusion broadens the homogeneous linewidth during the measurement timescale \cite{thiel_optical_2010} leading to a non-exponential echo decay often described by the empirical formula proposed by Mims \cite{mims_phase_1968,bottger_optical_2006}
\begin{equation}
I(t_{12})= I_0 \, \exp \left[ -2  \left( \frac{2 \, t_{12}} {T_{\rm M}} \right)^x \right] \; ,
\label{eq:echodecaymims}
\end{equation}
where the exponent $x \ge 1$ determines the decay shape and $T_{\rm M}$ is the phase memory lifetime, usually related to the effective homogeneous linewidth as $\Gamma_{\rm eff}=1/\pi T_{\rm M}$ \cite{thiel_optical_2010}. In heterodyne detection, the electric field of the emitted photon echo is detected rather than the intensity so that the decay of the echo heterodyne envelope area is slower by a factor of two and is given by
\begin{equation}
E(t_{12})= E_0 \, \exp \left[ - \left( \frac{2 \, t_{12}} {T_{\rm M}} \right)^x \right]  \; .
\label{eq:echodecayhd}
\end{equation}

\begin{figure}[t]
\centering
\includegraphics[width=1\columnwidth]{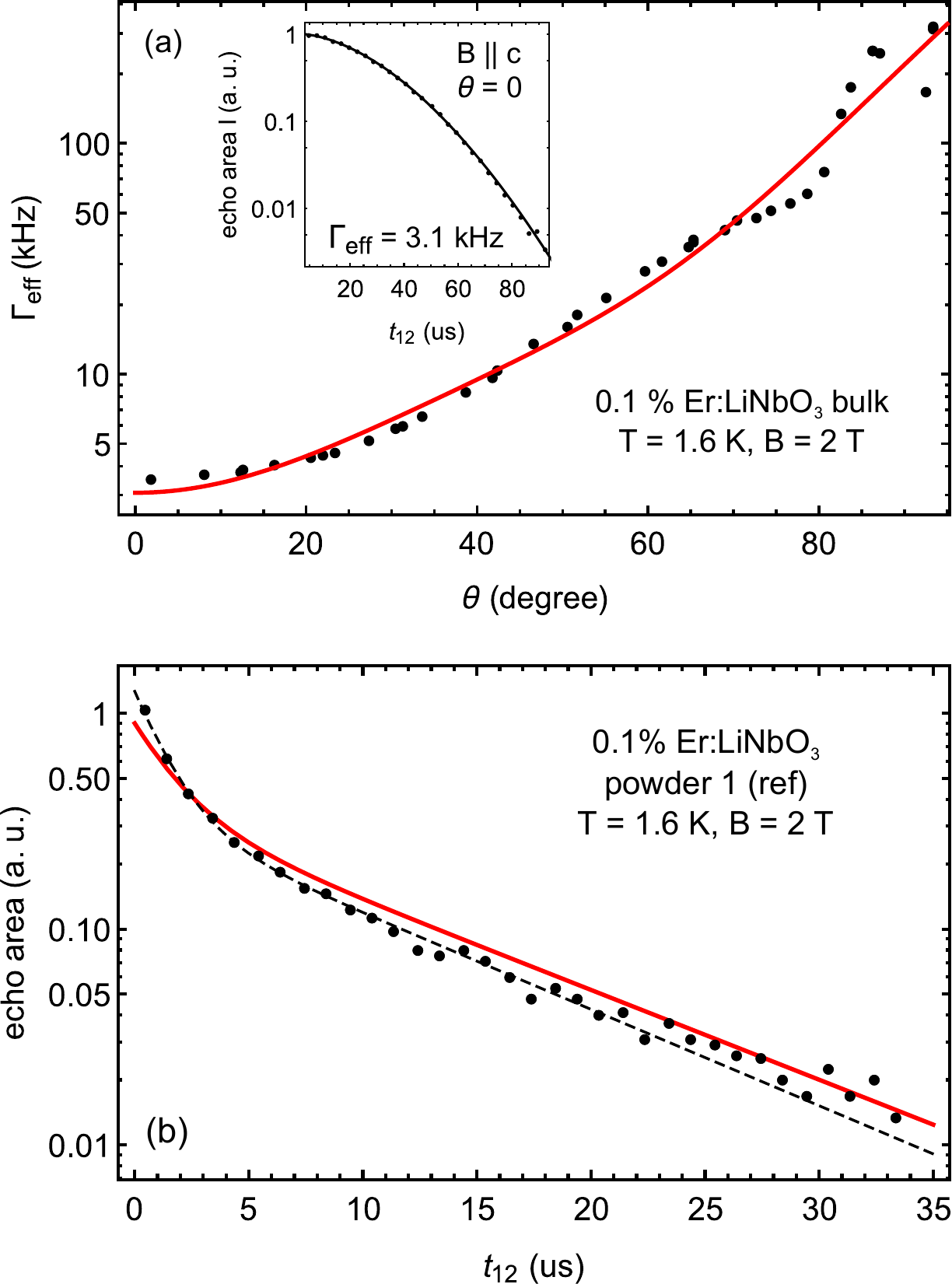}
\caption{(a) Orientation dependence of the effective linewidth $\Gamma_{\rm eff}$ in a 0.1\% \erlinbo bulk crystal measured at $T=1.6$~K and $B = 2$~T. The magnetic field orientation is rotated by an angle $\theta$ from the $c$-axis of the crystal. The inset shows an example echo intensity decay at $\theta=0$ i.e. $\mathbf{B} \parallel c$, where the fit of Eq.~\ref{eq:echodecaymims} gives $\Gamma_{\rm eff} = 1/\pi T_{\rm M}= 3.1$~kHz. The red solid line is an empirical polynomial fit to the experimental $\Gamma_{\rm eff}(\theta)$ data. (b) Decay of the echo area for the 0.1\% \erlinbo reference powder. The dashed line corresponds to a fit of a sum of two exponential decays to the experimental data (black dots). The red solid line corresponds to a calculation of the expected decay in the powder due to the orientation dependence of $\Gamma_{\rm eff}$ in the bulk crystal.}
\label{fig:echodecay}
\end{figure}

In general, the effective homogeneous linewidth can strongly depend on the orientation of the applied magnetic field $\mathbf{B}$ as well as the specific ion sites addressed in the crystal by the resonant optical excitation \cite{bottger_effects_2009}. For amorphous media, there can be a wide range of local environments and a corresponding need to average over a complex distribution of sites and orientations \cite{veissier_optical_2016}. In contrast, it should be possible to predict the behavior for a powder of randomly oriented crystallites from the measured orientation dependence of the bulk single crystal.
In LiNbO$_3$, Er$^{3+}$ ions substitute for Li$^+$ with a local environment that nominally maintains an axial point symmetry with respect to the $c$-axis described by the C$_3$ point group \cite{gog_x-ray_1993,garcia_sole_rare_1998}. Assuming that the sites maintain perfect axial symmetry, the orientation-dependent properties depend only on the angle $\theta$ between the magnetic field $\mathbf{B}$ and the $c$-axis. Fig.~\ref{fig:echodecay}(a) shows the effective linewidth as a function of $\theta$ in a 0.1\% \erlinbo bulk crystal at $T=1.6$ K and $B=2$ T. The minimum homogeneous linewidth is observed for $\mathbf{B}$ along the $c$-axis, i.e. $\theta =0$, with a measured value of 3 kHz, agreeing with the linewidth previously reported for these conditions in a 0.06\% \erlinbo crystal \cite{thiel_optical_2010}. The values of $\Gamma_{\rm eff}$ were extracted from 2PE intensity decays fitted by Eq.~\ref{eq:echodecaymims}, as shown in the inset for $\mathbf{B}$ along the $c$-axis. The measured angular dependence $\Gamma_{\rm eff}(\theta)$ was fit well using an empirical eighth order polynomial containing only even powers. To simulate the observed echo intensity $I_{\rm tot}$, we averaged the echo intensities over all possible random crystallite orientations as follows
\begin{equation}
I_{\rm tot}(t_{12})= I_0 \, \int_{0}^{\pi/2} \exp \left[-4 \pi \Gamma_{\rm eff}(\theta) t_{12}\right] \sin \theta \, d\theta \; .
\label{eq:intechodecay}
\end{equation}
This model assumes a uniform distribution of particle orientations and that all particles contribute equally to the averaged decay. Fig.~\ref{fig:echodecay}(b) shows the measured echo intensity as a function of waiting time $t_{12}$ in powder \#1 under the same conditions as for the bulk measurements, i.e. $B=2$ T and $T=1.6$ K, as well as the decay simulated using Eq.~\ref{eq:intechodecay} with the measured bulk crystal parameters. The simulated and measured decays agree very well, indicating that the observed echo signals in this case do represent a simple average over all possible particle orientations. This conclusion is important for interpreting the later results since it suggests that the probe light interacts uniformly with different particles and that there is no preferential alignment of the crystallites in the powder.

From both the simulations and the measurements, we observed echo decays for the \erlinbo powders that could be approximately described by a double-exponential shape. Using simulations to investigate the contributions of different particle orientations to the echo signal, we found that, for the conditions studied in this work, the slow component of the decay corresponds to the subset of crystallites in the powder that have their $c$-axis oriented almost parallel to the applied magnetic field direction while the fast component of the decay corresponds to all crystallites with other orientations. For example, under the conditions in Fig.\ref{fig:echodecay}(a), fitting the powder decay with a double exponential gives effective homogeneous linewidths of 7.6 and 45 kHz for the two components, roughly corresponding to that of the bulk crystal with $\mathbf{B}$ along the $c$-axis (3.5 kHz), and an average over the behavior for other orientations (up to 300 kHz  for $\mathbf{B}$ orthogonal to the $c$-axis), respectively.  

Because the slower decay component probes the subset of particles with their $c$-axis approximately aligned with the external field, as confirmed by the analytical model described by Eq.~\ref{eq:intechodecay}, we can quantitatively analyze the decoherence mechanisms by directly comparing this component of the decay to the bulk crystal behavior for $\mathbf{B} \parallel c$.  Consequently, in all of the following analysis, we focus on the slower component of the double exponential decay observed for the powder samples. In the spectral hole burning and free induction decay measurements described later, only single exponential decays are observed since at the timescales of these measurements, additional decoherence effects dominate the decay behavior and are more important than the orientation effect observed in the 2PE measurements.

\subsubsection{Magnetic field dependence of coherence lifetimes}

\begin{figure}[t]
\centering
\includegraphics[width=0.97\columnwidth]{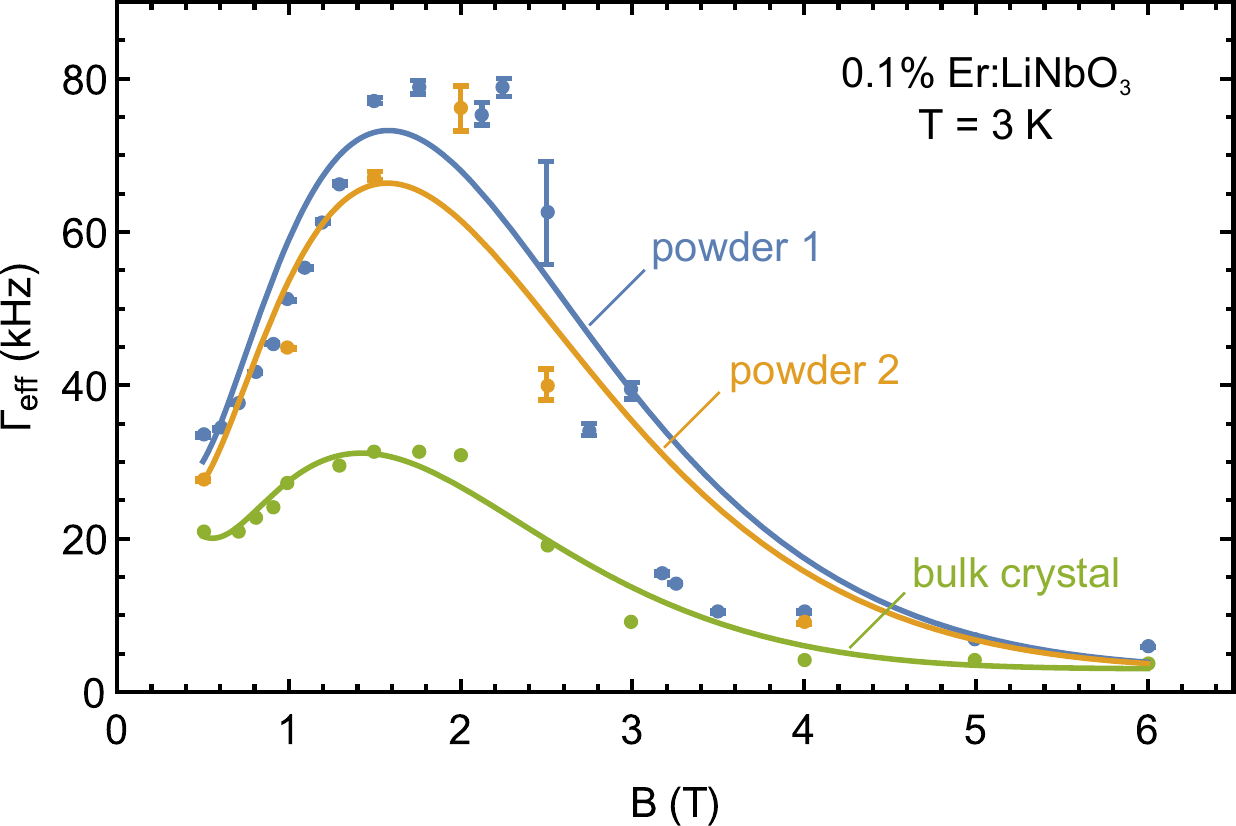}
\caption{Magnetic field dependence of the effective homogeneous linewidth in 0.1\% \erlinbo at $T=3$~K for the bulk crystal (for $\mathbf{B} \parallel c$), the reference powder \#1, and a powder ball-milled for 30 min. (powder \#2).}
\label{fig:Bdep3K}
\end{figure}

In this section we discuss the observed magnetic field dependence of the coherence lifetimes in the different samples and investigate additional decoherence processes resulting from the powder fabrication methods. Fig.~\ref{fig:Bdep3K} shows the magnetic field dependence of the effective linewidth at $T=3$ K for the bulk crystal, powder \#1 (the reference), and powder \#2. For all other powder samples listed in Tab.~\ref{tab:fab}, the signal was either too weak or exhibited a decay too rapid to extract a reliable lifetime from photon echo measurements. The experimental points $\Gamma_{\rm eff}(B)$ are fitted by the spectral diffusion model from Ref.~\cite{thiel_optical_2010}, where the effective linewidth is given by
\begin{equation}
\Gamma_{\rm eff}(B,T)=\frac{\Gamma_{\rm SD}R}{-2\Gamma_0+2\sqrt{\Gamma_0^2+\frac{\Gamma_{\rm SD}R}{\pi}}} \; ,
\label{eq:Gamma_eff}
\end{equation}
where $\Gamma_0$ is the linewidth without spectral diffusion, $\Gamma_{\rm SD}$ is the broadening caused by interactions between the Er$^{3+}$ electronic spins, and $R$ is the rate of spectral diffusion. The spectral diffusion broadening has a full width at half maximum described by
\begin{equation}
\Gamma_{\rm SD}(B,T)=\Gamma_{\rm max}\ \textrm{sech}^2\left(\frac{g_{\rm env}\mu_{\rm B} B}{2k_{\rm B} T} \right) \; ,
\label{eq:Gamma_SD}
\end{equation} 
where $\Gamma_{\rm max}$ is the maximum broadening due to magnetic dipole-dipole interactions, $g_{\rm env}\mu_{\rm B} B$ is the splitting of the Er$^{3+}$spin states, and the hyperbolic secant function in Eq.~\ref{eq:Gamma_SD} describes the suppression of spectral diffusion at high magnetic fields due to the increased magnetic ordering of the spins. Spectral diffusion occurs at a rate $R$ characterizing the flip rate of the perturbing Er$^{3+}$ spins; that is approximately given by
\begin{align}
R(B,T) = & \, \alpha_{\rm ff} \, \textrm{sech}^2\left(\frac{g_{\rm env}\mu_{\rm B} B}{2k_{\rm B} T} \right) \\ \nonumber
& + \alpha_{\rm D} \, g_{\rm env}^3 \, B^5 \textrm{coth}\left(\frac{g_{\rm env}\mu_{\rm B} B}{2k_{\rm B} T} \right) \; .
\label{eq:r_SD}
\end{align}
This expression assumes that only the mutual Er$^{3+}$-Er$^{3+}$ spin flip-flop process with coefficient $\alpha_{\rm ff}$ and the direct phonon process with coefficient $\alpha_{\rm D}$ drive Er\tplus spin flips, as expected for temperatures below 4 K.

\begin{table}[]
\centering
\begin{tabular}{|l|c|c|c|}
\hline
sample    & $g$ &  $\alpha_{\rm ff}$(kHz) &  $\alpha_{\rm_D}$(kHz/T$^5$) \\
\hline
bulk      		& 15.7     	& 11 &	69	\\
1	  	& 14      & 12 &	240	\\
2      	& 14       & 9	& 	200	\\
\hline
\end{tabular}
\caption{Parameters from fitting the experimental data shown in Fig.~\ref{fig:Bdep3K} by Eq.~\ref{eq:Gamma_eff}. The values of $\Gamma_{\rm max}$ was fixed to 1 MHz.}
\label{tab:fitparameters}
\end{table}

Fitting Eq.~\ref{eq:Gamma_eff} to the three sets of data (bulk crystal and powders \#1 and \#2) gives very similar spectral diffusion parameters except for the coefficient $\alpha_{\rm D}$ that is 4 times larger for powder \#1 and 3 times larger for powder \#2 as compared to the bulk crystal (see Tab.~\ref{tab:fitparameters}). This suggests that additional spin relaxation processes that increase with magnetic field are introduced through powder fabrication by crushing or ball-milling. Such processes may be understood within the framework of the two-level system (TLS) model developed to describe dynamics in amorphous materials \cite{Anderson1972,Phillips1972}. In this model, very low-frequency modes that involve groups of atoms tunneling between local configurations with nearly equivalent energies are enabled by defects and disorder in the crystal structure. It has been proposed that low densities of TLS can be present in some crystalline materials with large inhomogeneous lattice strains \cite{Watson1995}. Decoherence attributed to TLS has been previously observed in both bulk crystals \cite{Flinn1994,Macfarlane2004} and powders \cite{Meltzer1997}, and it has also been suggested that TLS can cause rapid electron spin relaxation \cite{Askew1986}. This interpretation is also consistent with past observations of increased electronic spin-lattice relaxation rates of Nd$^{3+}$ in YAG single crystals with a high density of structural defects in the lattice \cite{Aminov1998} as well as increased nuclear spin relaxation rates of $^{169}$Tm$^{3+}$ in YAG powders fabricated by ball milling \cite{lutz_effects_2016}. While it is known that high-energy ball milling of nanocrystals can produce significant disorder in the crystal structure and introduce some amorphous behavior \cite{bork_nmr_1998,pooley_synthesis_2003,heitjans_fast_2004,chadwick_lithium_2005,heitjans_nmr_2007}, our results suggest that even very low-energy grinding methods applied to much larger microcrystals still cause substantial damage that produce TLS.

Because 2PE techniques did not allow us to measure coherence lifetimes in all of the powder samples, we also used SHB and FID techniques to probe the effects of faster decoherence processes, as described in the following sections.

\subsection{Spectral hole burning}

We employed spectral hole burning to measure $\Gamma_{\rm eff}$ as well as the effective absorption depth $\alpha_{\rm eff}$ of our samples. We use the latter  to qualitatively estimate the amount of scattering and the quality of the micro/nano-crystals. In this measurement, a 500 $\mu$s long burn pulse excited the resonant Er$^{3+}$ ions into the $^4$I$_{13/2}$ excited state and created a corresponding transparency window in the inhomogeneously broadened absorption line. After a 25 $\mu$s delay, a subsequent 50 $\mu$s long read pulse with lower power was frequency swept across the spectral hole to measure the absorption spectrum.

\begin{figure}[t]
\centering
\includegraphics[width=1\columnwidth]{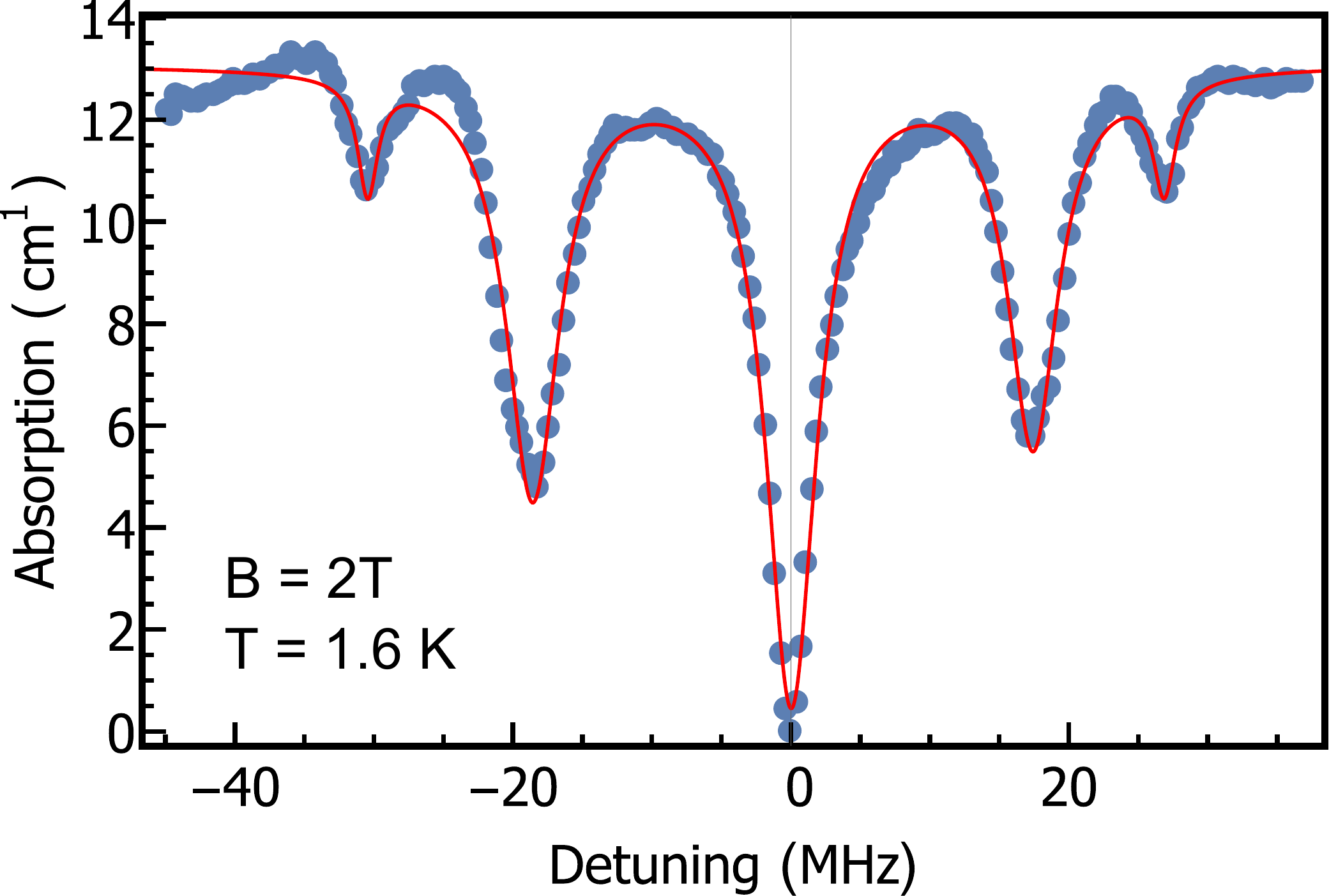}
\caption{Hole burning spectrum in 0.1\% \erlinbo (powder \#1) at $T = 1.6$~K and $B = 2$~T. Zero detuning corresponds to 1531.882 nm in vacuum. }
\label{fig:hb}
\end{figure}

A typical hole burning trace is shown in Fig.~\ref{fig:hb} for powder \#1 at a temperature of 1.6 K and a magnetic field of 2 T. In addition to the main hole in the center of the spectrum, we observed additional sideholes originating from the superhyperfine coupling between the Er$^{3+}$ electronic spins and the nuclear spins of the host ions, with $^{93}$Nb producing a superhyperfine splitting of 9 MHz/T and $^7$Li producing a splitting of 14.8 MHz/T \cite{Thiel2012}.

Spectral hole burning measurements provide information about the optical coherence lifetime via the width of the spectral hole. 
In the absence of power broadening, the spectral hole width is twice the effective homogeneous linewidth convolved with the laser linewidth \cite{coherent_1987}. To estimate the effect of power broadening and the laser linewidth, we carried out a series of hole burning measurements on the bulk crystal using different laser powers. This measurement allowed us to extrapolate to the hole width at zero burning power. From a power broadening measurement performed on powder \#3, we estimated that our measured hole widths exhibited 1.93 MHz of power broadening, and we then assumed that the power broadening was the same for each powder under the same excitation conditions. This gives a large uncertainty that we estimated to be 10 kHz on the extracted values of the effective linewidth, but still allows us to compare the effects of the different treatments applied to the powders. A comparison between SHB and 2PE measurements for the bulk crystal resulted in an estimated contribution of the laser linewidth to the hole width of 420 kHz. Using these values, the effective homogeneous linewidths were computed from the hole burning measurements by subtracting the power broadening and laser linewidth contributions, with the outcome summarized in Tab.~\ref{tab:hb}.


\begin{table}[t]
\centering
\begin{tabular}{|l| c| c| c|c|}
\hline
sample  & $\alpha_{\rm eff}$ (cm$^{-1}$) & $\Gamma_{\rm eff}^{\rm HB}$ (kHz) &  $\Gamma_{\rm eff}^{\rm 2PE}$ (kHz) &  $\Gamma_{\rm eff}^{\rm FID}$ (kHz)\\
\hline
bulk   	 & 3.9	& $<$ 10	&3.2 		&	$<$10	\\
1 (ref)  & 13.5	& 60   		&4.5/40 	&	32	\\
3     	 & 9.4	& $< 10$  	&12/45		&	38	\\
4        & 1.4	& 1050 		&30 		&	30	\\
5        & 0.71	& 990  		&38			& 	29	\\
6        & 2.96	& 150  		&37			&	42	\\ 
\hline
\end{tabular}
\caption{Maximum effective absorption coefficient $\alpha_{\rm eff}$ extracted from SHB measurements, as well as effective linewidths $\Gamma_{\rm eff}$ extracted from SHB, 2PE and FID (after a t$_w =$ 5 $\mu$s) measurements at $T=1.6$~K and $B=2$~T for different samples.  When a double exponential decay was observed, as described in the main text, both decay values are listed. For the bulk crystal, the magnetic field and light polarization were parallel to the $c$-axis ($\mathbf{B} \parallel \mathbf{E} \parallel c$).}
\label{tab:hb}
\end{table}

From the values in Tab.~\ref{tab:hb}, we see that powder \#1, which was crushed using only a mortar and pestle, shows broader hole widths than the bulk crystal, meaning that additional spectral diffusion processes occur in this powder. Remarkably, the effective homogeneous linewidth of powder \#3, which was ball-milled and then annealed, is narrower, and on the same order of magnitude as that of the bulk crystal. This indicates that crystal damage (strain and defects) caused by mechanical processing, in this case crushing using a mortar and pestle and slow ball milling, could be mostly removed by annealing the powder with the smallest particle sizes studied of roughly 2 $\mu$m and lower. On the other hand, powder \#4 shows much broader spectral holes with the same mechanical processing as powder \#3 but annealing at a higher temperature. To further explore the effect of temperature, powder \#5 was ball milled at a higher speed and for a longer time than powders \#3 and \#4, but was then annealed at the same upper temperature as powder \#4, resulting in very large spectral hole widths -- nearly the same as for powder \#4. These results suggest that the much broader linewidths observed in powders \#4 and \#5 arise from the higher annealing temperatures. Consequently, it appears that temperatures above 1000~C are detrimental to the material properties rather than beneficial, likely due to additional disorder caused by loss of either lithium or oxygen from the crystal matrix.  This hypothesis is consistent with past observations of substantial lithium out-diffusion from the surface of bulk crystals when annealed in a wet oxygen atmosphere at temperatures above approximately 1100~C \cite{forouhar_effects_1984}. Moreover, changes in material properties due to lithium out-diffusion have been observed at temperatures as low as 600~C for LiNbO$_3$ thin films \cite{wang_influence_2007,budakoti_enhancement_2008}, suggesting that even lower annealing temperatures than used here may be optimal for micro/nano-powders.

Finally, we examined powder \#6 that was fabricated through in-diffusion of erbium into commercial LiNbO$_3$ powder. This powder was processed using similar mechanical and thermal treatments as for powders \#4 and \#5, although noticeably narrower linewidths than those of powders \#4 and \#5 were observed. Further study is required to determine whether the difference in linewidth is a result of the different starting materials, the presence of ErCl$_3$ mixed with the powder, or other factors.

From the hole burning spectra, we also extracted effective absorption depths for each powder sample. Since scattering in the powder increases the optical path length traveled by the light as it passes through the sample, we should observe a larger absorption depth for samples with more scattering, as well as the expected increase in the overall frequency-independent scattering loss. On the other hand, scattering of light between and around loosely packed particles will reduce the absorption compared to a bulk crystal. In particular, any small gap in the 0.4~mm thick loose powder can result in a weakly scattered background signal that does not strongly probe the powder. Furthermore, when the individual crystallites have anisotropic absorption, the transmitted light in general will not follow the usual Beer-Lambert-Bouguer exponential decay law.  Nevertheless, if we assume that the incident light polarization is randomized many times during the scattering process, as we observe experimentally from the transmitted light, then we can approximate the propagation through the powder using the absorption coefficient averaged over all polarizations states. Thus, while it is difficult to extract real absorption coefficients or effective path lengths from the measured transmission spectrum, we do expect the observed “effective” absorption to provide some qualitative information about the relative scattering path lengths in the powder samples.

For the 1532 nm transition of \erlinbo, the absorption coefficient is dependent on both the polarization and the propagation direction of the light since the transition has comparable electric and magnetic transition dipole strengths \cite{Thiel2012}. The maximum absorption coefficients are known for light propagating orthogonally to the crystal\textquotesingle s $c$-axis with $\pi$ polarization, i.e. $\mathbf{E} \parallel c$ (4.1 cm$^{-1}$), and $\sigma$ polarization, i.e. $\mathbf{E} \perp c$ (6.8 cm$^{-1}$), as well as for the light propagating along the crystal\textquotesingle s $c$-axis with the two circular polarization states (11.1 cm$^{-1}$ averaged over left and right handed polarizations). Consequently, averaging over all possible random crystal orientations and polarizations, we would expect an effective absorption coefficient $\alpha_{\rm eff}$ of 7.3 cm$^{-1}$ for a powder in the limit of many scattering events.

To estimate the “effective” absorption coefficient $\alpha_{\rm eff}$ for each powder sample, we used high-power (resonant) optical excitation to burn a very deep spectral hole for which we assumed complete transparency since further increasing the burning power did not produce any additional increase in the transmission. Using this approach, $\alpha_{\rm eff}$ was calculated from the transmitted intensity $I_t$ at the center of the hole and the intensity $I_0$ with the laser tuned away from the hole (where it experiences the full absorption) through the relation
\begin{equation}
\alpha_{\rm eff} = -  \ln \left( \frac{I_t}{I_0} \right) /L \; ,
\label{eq:a_eff}
\end{equation}
where $L = 0.4$~mm is the powder thickness.  The resulting effective absorption values are listed in Tab.~\ref{tab:hb} for each sample.

For powders \#1 and \#3 we measured absorption coefficients (see Tab.~\ref{tab:hb}) higher than the bulk average increased by a factor of 1.8 and 1.3 respectively. This is consistent with the transmitted light experiencing a longer path in the powder sample. 
However, for the other powder samples, the measured absorption coefficient is significantly lower compared to the bulk. We note that all powders  with small optical depth (\#4 to \#6)  underwent the same annealing procedure, with maximum temperatures of around 1100~C, which might have introduced modifications in the host matrix that potentially
could cause broadening of the inhomogeneous linewidth or other effects responsible for
decreased absorption. This explanation would also be consistent with our observation of larger effective homogeneous linewidths in those same samples.  Nevertheless, to conclusively identify the cause of the weaker observed spectral hole depths in these samples requires further study.

The experiment described above shows that SHB measurements in direct transmission could be successfully conducted in powder samples and that the effective linewidths were successfully extracted in samples for which 2PE measurements could not be performed. Overall, we observed that coherence properties degraded in powders that underwent mechanical processing. This suggests the presence of two-level systems in these samples. To further verify these results, we also used FID techniques to probe the coherence properties over an intermediate range that overlaps those accessible through SHB and 2PE measurements.

\subsection{Free induction decay and spectral diffusion}

SHB measurements may be used to probe powders with broad homogeneous linewidths that cannot be accessed by 2PE techniques. However, in cases where rapid spectral diffusion or complex spectral structure such as spin-flip sidebands are present, optical coherent transient FID measurements can serve as a better method to extract the effective homogeneous linewidth \cite{coherent_1987}. While FID measurements that probe a single spectral hole are still limited by the laser linewidth, FID may be employed immediately following the hole preparation without the need for laser chirping, allowing the broadening to be probed over shorter timescales than frequency domain SHB measurements. Furthermore, FID’s are more amenable to heterodyne detection techniques, allowing for improved signal discrimination and dynamic range relative to SHB. The shape of the FID curve also reveals the evolution of decoherence processes over the timescale of the measurement in a manner similar to 2PE techniques \cite{thiel_optical_2010}, unlike direct SHB that requires an extensive series of measurements to extract similar information.

\begin{figure}[t]
\centering
\includegraphics[width=1\columnwidth]{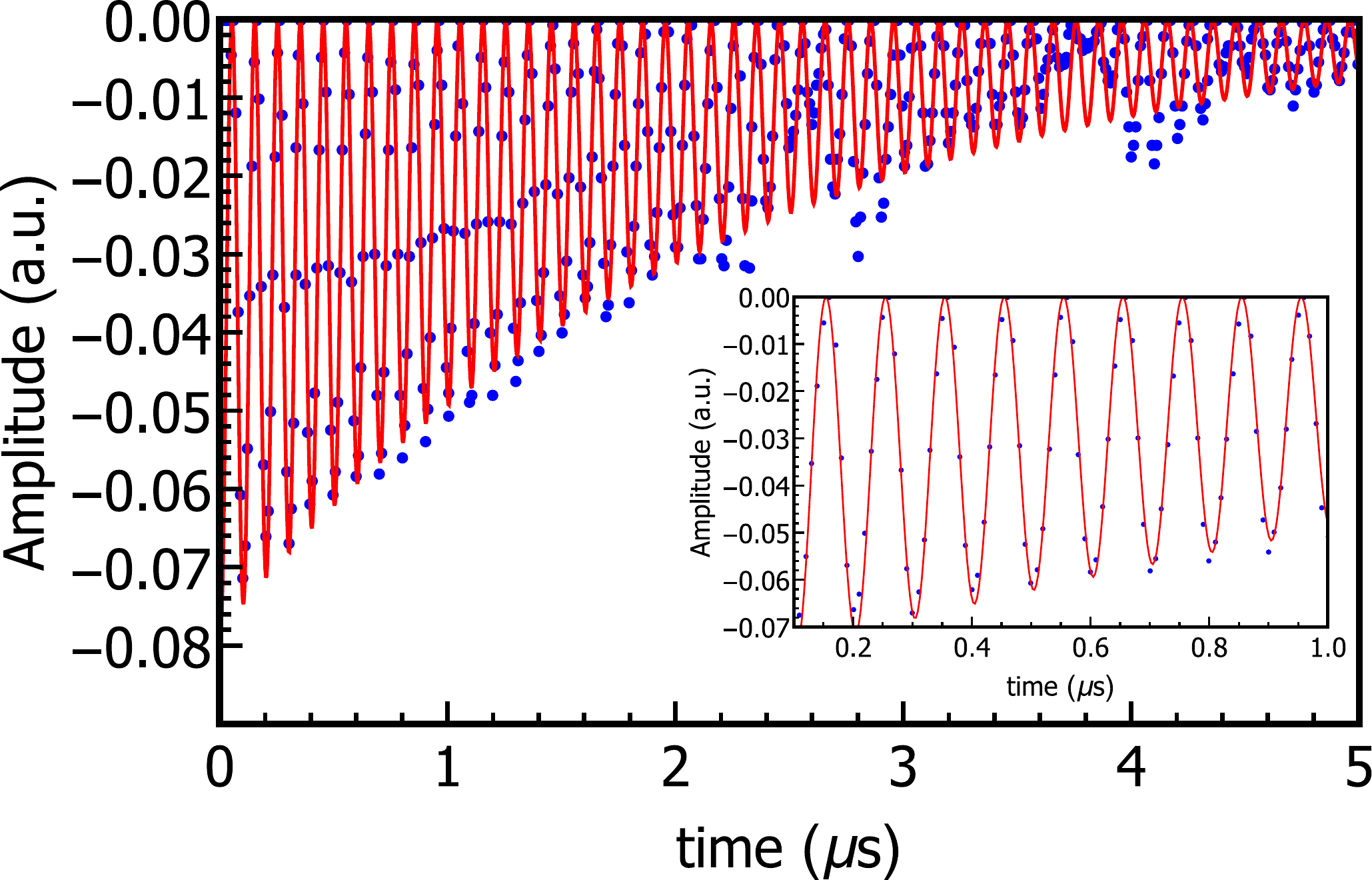}
\caption{Typical background-subtracted FID decay of the reference powder \#1 at $B=3.9$~T, $T=1.6$~K and waiting time $t_{\rm w} = 5$~$\mu$s. Solid points are measured values and the solid line is a fit of the FID signal. The inset shows a magnification of the first 1 $\mu$s of the decay.}
\label{fig:fiddec}
\end{figure}

Motivated by the ability of FID measurements to probe rapid decoherence processes contributing to the homogeneous linewidth (i.e. processes faster than the shortest timescales observable using 2PEs), we employed FID measurements to study all of our powder samples. Specifically, we applied a variation of the FID technique sometimes referred to as a “delayed” FID that is analogous to a stimulated photon echo measurement \cite{coherent_1987}. For this approach, a laser pulse first burnt a hole in the inhomogeneously broadened absorption line of the sample: in our case we used a 50 $\mu$s long burn pulse to provide an approximately 20 kHz Fourier-limited spectral resolution. After a waiting time $T_{\rm w}$, a second brief excitation pulse was applied to induce an optical coherence across the frequency window of the burnt hole. The decay of the coherent emission after the brief pulse provides information about the effective width of the spectral hole resulting from the convolution of the effective homogeneous linewidth and the laser linewidth, as well as any spectral diffusion that occurs over the waiting time $T_{\rm w}$. In our case, we employed heterodyne detection to measure the coherence decay by beating it with a third laser pulse shifted in frequency by 10 MHz. The decay of the resulting 10 MHz beat signal is given by
\begin{equation}
E(t)= E_0 \, \exp \left[ - \left( \frac{t} {T_{\rm M}^{T_{\rm w}}} \right)^x \right]  \; .
\label{eq:fiddec}
\end{equation}
By fitting this equation to the experimentally measured decay, we extracted the effective homogeneous linewidth $\Gamma_{\rm{eff}}=1/\pi T_{\rm M}^{T_{\rm w}}$ after a waiting time $T_{\rm w}$. An exemplary FID decay signal for powder \#1 (reference) at $B=3.9$~T, $T=1.6$~K and waiting time $T_{\rm w}=5$ $\mu$s is shown in Fig.~\ref{fig:fiddec}, where the steady-state transmitted intensity has been subtracted from the signal. The fit of the FID is shown by the solid line, revealing an exponential decay envelope (i.e. $x = 1$ in Eq.~\ref{eq:fiddec}) that corresponds to an effective linewidth of $\Gamma_{\rm eff} = 147$~kHz.

\begin{figure}[t]
\centering
\includegraphics[width=1\columnwidth]{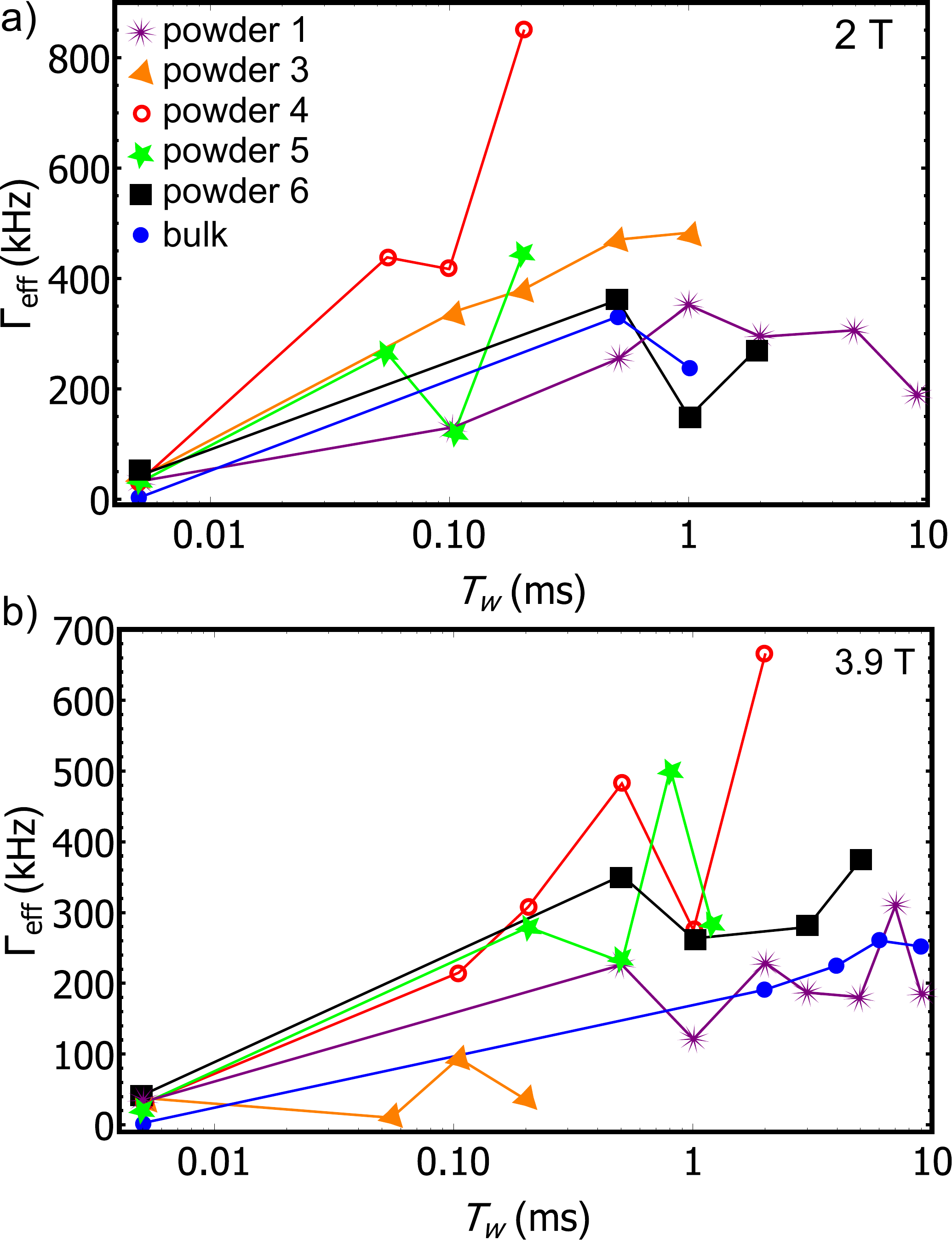}
\caption{Effective homogeneous linewidths at a) $B = 2$~T and b) $B = 3.9$~T at $T = 1.6$~K as a function of the waiting time $T_{\rm w}$ measured using the delayed FID
technique for powders \#1, \# 3 through \#6 and for the bulk crystal. The solid lines are guides to the
eye.}
\label{fig:fiddiff}
\end{figure}

By changing the waiting time $T_{\rm w}$ between the burn and excitation pulses, the delayed FID technique was used to determine the effective homogeneous linewidth at different timescales and thus study time-dependent decoherence processes, i.e. spectral diffusion. The effective homogeneous linewidth over a range of waiting times for different powders is shown in Fig.~\ref{fig:fiddiff}. From a comparison of the obtained effective homogeneous linewidths using FID at $T_{\rm w}=5$ $\mu$s and the local slope of the two pulse photon echo
decay around $t_{12}=5$ $\mu$s for each powder, we estimated the experimental contribution to each FID measurement due to laser linewidth, power broadening and shot-to-shot laser jitter.

As shown in Fig.~\ref{fig:fiddiff}, we observed a rapid time-dependent broadening of the linewidth due
to spectral diffusion occurring at timescales of less than 500 $\mu$s, and then a saturation of the broadening at longer timescales, indicating that those waiting times were longer than the inverse of the spectral diffusion rate. Comparing the behaviors at $B=2$~T and $B=3.9$~T (see Fig.~\ref{fig:fiddiff}(a) and (b)) indicates that spectral diffusion is faster at 2 T, whereas the final linewidth at long $T_{\rm w}$ is similar at both magnetic fields. Overall, the bulk crystal shows the narrowest linewidths, although powder \#1 (the reference powder) that underwent minimal grinding using only a mortar and pestle exhibits similar spectral diffusion broadening as the bulk crystal at the longest time delays. All ball-milled powders show significantly larger effective linewidths. For powders \#3 and \#4 that underwent the same mechanical treatment,
the data points at short $T_{\rm w}$ suggest that the higher annealing temperature of 1100~C for powder \#4 was detrimental. Furthermore, we could not
observe a clear saturation of the effective homogeneous linewidth with waiting time for powder \#4, unlike
all of the other powders. Powder \#5 that experienced more severe mechanical treatment
(fast ball milling for 10 hours) but the same annealing process as \#4, shows comparable
broadening as \#4, further suggesting that the high annealing temperature was the limiting factor for these samples. This observation is consistent with the SHB results discussed earlier and again points toward defects and disorder in the crystal caused by out-diffusion of lithium at high temperatures \cite{forouhar_effects_1984,wang_influence_2007,budakoti_enhancement_2008}. Since the dynamic TLS modes enabled by the disorder typically exhibit a broad distribution of relaxation rates, they are expected to cause significant spectral diffusion that often produces a logarithmic increase in the homogeneous linewidth over time \cite{black_spectral_1977,silbey_time_1996}.

The commercial powder with large crystallites that were diffusion-doped with Er$^{3+}$ (powder \#6) exhibited better coherence properties at long timescales compared to all of the ball-milled powders, although still significantly worse than the bulk crystal and the crushed reference powder \#1. This again suggests that higher processing temperatures are a key factor in producing spectral diffusion, although at a lower level for larger particles as might be expected if lithium out-diffusion is the cause. 

\section{Conclusion}

We demonstrated that spectroscopic properties such as optical coherence lifetimes in \erlinbo powders that experienced various mechanical and thermal treatments could be measured despite strong scattering, random orientations, and shorter coherence lifetimes. The 2PE measurements were optimal to access long coherence lifetimes whereas SHB and FID techniques were well suited to quantify fast dephasing processes. Furthermore, we described how the powder properties could be predicted and understood by averaging the orientation dependent properties of the bulk single crystal.

We observed that mechanical grinding of the powders resulted in a broadening of their effective homogeneous linewidths, indicating the presence of dynamic TLS modes characteristic of amorphous phases. This conclusion was also supported by observations of corresponding decreases in the maximum absorption as well as a broader distribution of spectral diffusion rates as compared to the bulk crystal.
Surprisingly, we found that even crushing and grinding with mortar and pestle to produce large microcrystals resulted in significant damage as indicated by an order-of-magnitude increase in the homogeneous linewidth. In the powders that experienced less processing, spectral diffusion broadening at longer timescales approached the values observed in the bulk, suggesting that grinding primarily accelerates existing spectral diffusion mechanisms (such as Er$^{3+}$ electronic spin flips) rather than introducing new sources of broadening. This conclusion is also consistent with the increase in the magnetic-field‐dependent spectral diffusion rates extracted from analysis of the 2PE coherence measurements (summarized in Tab.~\ref{tab:fitparameters}). We found that the broadening could be partially reversed by high-temperature annealing in oxygen atmospheres for some powders; however, treatment at temperatures above 1000 C was detrimental for the coherence properties. The powders annealed at temperatures above 1000 C exhibited a large increase in the magnitude of spectral diffusion, suggesting that additional broadening mechanisms were introduced, possibly due to out-diffusion of lithium and oxygen from the crystallites.

Together, our results demonstrate that even minimal low-energy mechanical processing significantly degrades the material properties of interest for classical and quantum signal-processing applications. Chemical fabrication methods, such as chemical synthesis or etching may minimize the amount of amorphous-like behavior introduced when fabricating micro- and nano-materials and lead to better optical coherence properties. This hypothesis is consistent with past structural and spectroscopic studies of nanocrystalline LiNbO$_3$ where powders produced by sol-gel methods had significantly better properties than those produced by high-energy ball milling \cite{pooley_synthesis_2003,chadwick_lithium_2005}, although still not reaching the bulk crystal properties \cite{heitjans_nmr_2007}. The detrimental effects of high annealing temperatures also suggest that lower temperature bottom-up synthesis may be preferred. The demonstration of long coherence lifetimes in Eu:Y$_2$O$_3$ nanopowders produced using solvothermal methods confirms that at least some materials can be fabricated with optical coherence properties approaching those of the bulk crystal \cite{perrot_narrow_2013}. Based on the recent work on Nd:Y$_2$SiO$_5$ \cite{zhong_nanophotonic_2015}, we expect that focused-ion-beam milling may also be successful in preserving the coherence properties of REI-doped LiNbO$_3$, potentially providing a suitable tool to fabricate micro- and nano-structures for optical signal-processing applications. Nevertheless, further studies are required to determine whether there are other fundamental limitations of material performance when the crystal size is reduced to the nanoscale.

\section{Acknowlegments}

The authors wish to thank Ph. Goldner and A. Ferrier for valuable discussions on optical coherence measurements in powders. The work reported here was supported by Alberta Innovates Technology Futures (ATIF), the National Engineering and Research Council of Canada (NSERC), and the National Science Foundation of the USA (NSF) under award nos. PHY-1212462, PHY-1415628, and CHE-1416454. W. T. is a senior fellow of the Canadian Institute for Advance Research (CIFAR).

\end{document}